\documentclass[letterpaper,11pt]{article}
\usepackage{jheppub} 
\usepackage{graphicx}
\usepackage{amsmath,amsthm, amssymb}
\usepackage{braket}
\usepackage{natbib}
\usepackage{xcolor}

\newcommand{\R}{\mathbb{R}}
\newcommand{\C}{\mathbb{C}}

\newcommand{\N}{\mathcal{N}}
\newcommand{\coset}{H_3^+}
\newcommand{\D}{\partial}
\newcommand{\dbar}{\bar \partial}
\newcommand{\p}{\psi_+}
\newcommand{\m}{\psi_-}
\newcommand{\ptild}{\tilde \psi_+}
\newcommand{\mtild}{\tilde \psi_-}
\newcommand{\lp}{\lambda_+}
\newcommand{\lm}{\lambda_-}
\newcommand{\lptild}{\tilde \lambda_+}
\newcommand{\lmtild}{\tilde \lambda_-}
\newcommand{\V}{\mathcal{V}}
\newcommand\normalorder[1]{\mathopen{:}#1\mathclose{:}}

\title{A supersymmetric AdS$_3$ duality}
\author{Bhanu Narra}
\affiliation{Pembroke College, University of Oxford, \\ St. Aldates, Oxford, OX1 1DW, UK }
\emailAdd{bhanu.narra@pmb.ox.ac.uk}

\abstract{We find a dual description of the supersymmetric $SL(2, \mathbb{C})/SU(2)$ WZW model as a super-winding condensate CFT, which follows simply from the bosonic AdS$_3$ duality found in [arXiv:2104.07233]. This duality can be seen as a dimensional uplift of the mirror symmetry between the $H_3^+/U(1)$ supercoset and $\mathcal{N}=2$ super-Liouville theory.}

\begin{document}

\maketitle

\section{Introduction}

In this note, we find the supersymmetric analogue of the bosonic (E)AdS$_3$ duality found in \cite{ER_EPR, Halder_2023}. On one side of the bosonic AdS$_3$ duality is the $SL(2, \C)/SU(2)\simeq \coset$ WZW model, which is the worldsheet theory of a bosonic string in EAdS$_3$. The dual is a winding condensate CFT given by a linear dilaton +  $\beta\gamma$ system deformed by $(1, 1)$ winding operators 
\begin{equation}\label{bosonic potential}
    W_\pm = e^{-\sqrt{k_B-2}\rho}e^{\frac {\pm\sqrt{k_B}} 4 (\gamma+ \bar \gamma )}e^{\mp\sqrt{k_B}(\int^z \beta(z')dz'+ \int^{\bar z} \bar \beta(\bar z')d\bar z')}.
\end{equation}
EAdS$_3$ can be visualized as a solid cylinder with time running along its length. The dual geometry is a cylinder with its core removed, meaning spatial slices are annuli instead of disks. Winding strings wrap this core, giving rise to the potential (\ref{bosonic potential}). 

The AdS$_3$ duality is a dimensional uplift of the FZZ duality, which relates the $\coset/U(1)$ coset CFT to sine-Liouville theory \cite{Kazakov_2002}. The dualities are related as follows: both the $\coset$ WZW model and the winding condensate CFT have a current $J_3$ which represents a timelike isometry within spacetime. Gauging this isometry in the WZW model gives the coset CFT; gauging it in the winding condensate gives sine-Liouville theory. 

The FZZ duality has an $\mathcal{N}=2$ supersymmetric analogue, which relates the $\coset/U(1)$ Kazama-Suzuki supercoset theory to $\mathcal{N} =2 $ super-Liouville theory \cite{Giveon_1999, SUSY_FZZ}. In this case, the two theories are related by mirror symmetry \cite{SUSY_FZZ}. It is natural to seek a corresponding supersymmetric analogue of the AdS$_3$ duality, which would relate the supersymmetric $\coset$ WZW model to some super-winding condensate CFT.\footnote{Note that the $\coset$ super-WZW model only has $\mathcal{N}=1$ supersymmetry. It is enhanced to $\mathcal{N}=2$ upon gauging $U(1)$ \cite{KAZAMA1989232}.} 

We will show that the supersymmetric AdS$_3$ duality follows trivially from the bosonic case due to the following fact: the supersymmetric $\coset$ WZW model at level $k$ factorizes into a bosonic WZW model at level $k_B = k+2$ and three free fermions. We can therefore simply replace the bosonic part of the super-WZW model with its dual winding condensate CFT whilst leaving the fermions unchanged. Hence, the dual to the supersymmetric theory is obtained by simply adding three free fermions to the bosonic winding condensate CFT! The explicit map between normalizable vertex operators and the equivalence of their two- and three-point correlators follows immediately from \cite{Halder_2023} due to the factorization of the theory. To verify that this is indeed the intended dual to the super-WZW model, in the sense that it is a `dimensional uplift' of the supersymmetric FZZ duality, we show that this super-winding condensate CFT reduces to $\mathcal{N}=2$ Liouville theory upon gauging the timelike isometry $J_3$, in analogy with the bosonic case. 

In section \ref{FZZ review}, we briefly review some relevant details of the 2D supersymmetric FZZ duality. In section \ref{uplift}, we describe how the 3D supersymmetric duality follows from the bosonic case and argue that the potential reduces to the Liouville potential.

\section{Review of the supersymmetric FZZ Duality}\label{FZZ review}

\subsection{The supersymmetric EAdS$_3$ WZW Model}

EAdS$_3$ is equivalent to the hyperbolic upper half-plane $\coset$, which is the coset $SL(2, \C)/SU(2) $. This is the space of complex hermitian $2\times2$ matrices with determinant one. A general element $h\in \coset$ can be parameterized by radial coordinate $r\in[0, \infty)$, angular coordinate $ \theta \sim \theta+2\pi$, and (Euclidean) temporal coordinate $ \xi\in(-\infty, \infty)$ as follows:
\begin{equation}
    h  = \begin{pmatrix}
        \cosh(r)e^\xi & \sinh(r)e^{i\theta} \\ \sinh(r)e^{-i\theta} & \cosh(r)e^{-\xi}
    \end{pmatrix}.
\end{equation}
The metric in these coordinates takes the form 
\begin{equation}\label{AdS metric}
    ds^2 = l^2(dr^2 + \sinh^2(r)d\theta^2 + \cosh^2(r)d\xi^2),
\end{equation}
where $l$ is the AdS$_3$ radius in string units.\footnote{We will set $\alpha'=1$.}

The action can be obtained by continuing the supersymmetric $SL(2, \R)$ WZW model at level $k = l^2$. It can also be found as the supersymmetric non-linear sigma model into the metric (\ref{AdS metric}) with a background $B$-field $B=2il^2\sinh^2(r)d\theta \wedge d\xi$,\footnote{An exact term $d(\ln(\cosh(r)d\theta)$ in the $B$-field has been dropped here.} which comes from the WZ term.

One can decouple the fermions within the SUSY WZW action via a field redefinition. This leaves a bosonic WZW model into $\coset$ at level $k_B=k+2$ along with three free fermions. The adjustment to the level comes from a chiral anomaly in the path integral measure \cite{DIVECCHIA1985701, Schnitzer:1988qj}. Pairing two of the fermions as $\psi_\pm = \frac 1 {\sqrt{2}}( \psi_1\pm i\psi_2), \tilde \psi_\pm =\frac 1{\sqrt{2}} (\tilde \psi_1\pm i \tilde \psi_2)$, the action takes the form 
\begin{align}\label{SUSY WZW Action}
    S_{\text{WZW}}= \frac{(k+2)}{2\pi} &\int d^2 z \left[\partial r \bar \partial r + \partial \xi \bar \partial \xi +(\partial \xi - i\partial \theta )(\bar \partial \xi + i \bar \partial \theta) \sinh^2(r) \right] \nonumber \\+\frac{1}{2\pi} &\int d^2z \,  \left[\psi_+\dbar \psi_-+\tilde\psi_+\D \mtild +\frac 1 2 \left(\psi_3\dbar \psi_3+\tilde \psi_3\D \tilde \psi_3\right) \right].
\end{align}

We will later be interested in the asymptotic form of this action as $r\to \infty$. The fermionic part is unaffected. In order to avoid the divergence of $\sinh^2(r)$ in the bosonic part of the action, we can introduce the first-order variables $W:=\xi+i\theta$ and an auxilliary field $\chi$ \cite{ER_EPR}:
\begin{equation}
    S_{\text{WZW}}=\frac{(k+2)}{2\pi}\int d^2z\left[\D r \dbar r + \partial \xi \bar \partial \xi + \chi  \dbar W +\bar \chi \D \bar W -\frac 1 {\sinh^2(r)}\chi \bar \chi\right] + S_\text{fermi}.
\end{equation}
This change of variables gives rise to a dilaton profile $\Phi(r) \sim -\ln(\sinh(r))$ in the quantum theory. One can reobtain the action (\ref{SUSY WZW Action}) by integrating out $\chi$. If we define the normalized fields\footnote{The differing normalization of $\rho$ comes from quantum corrections to the action; see \cite{ER_EPR}.}
\begin{equation}
    \rho  := \sqrt{k} r, \; \gamma : = \sqrt{k+2}W, \; \beta : = \sqrt{k+2} \left(\frac 1 2 \D \xi + \chi\right) 
\end{equation}
and take $\rho\to \infty$, this simply becomes 
\begin{equation}
    S_\text{free}=\frac{1}{2\pi}\int d^2z\left[\D \rho \dbar \rho + \beta  \dbar \gamma +\bar \beta \D \bar \gamma \right] + S_\text{fermi}.
\end{equation}
The dilaton is asymptotically linear, $\Phi(\rho) \to -\rho/\sqrt{k} $. The scalar $\rho$ therefore has a background charge $Q=1/\sqrt{k}$ in the asymptotic CFT. The $\beta \gamma$ system has weights $[\beta] =1 , [\gamma] = 0$. Note that $\text{Im} [\gamma] = \sqrt{k+2} \theta$ has periodicity $2\pi \sqrt{k+2}$.

\subsubsection{Gauging $U(1)$}\label{gauging}

In order to compare with the supersymmetric FZZ duality, we must gauge the $U(1)$ timelike isometry of $\coset$, yielding the Kazama-Suzuki supercoset theory into $\coset/U(1)$. Semiclassically, this coset theory is a sigma model into a 2D cigar. To see this, we write the action as a gauged WZW model \cite{Nakatsu:1991pu, Schnitzer:1988qj, Witten:1991mk}. One must replace $\partial \xi \to \partial \xi +A$ in the bosonic part, replace $\partial \to \partial +A$ in the fermionic part, and set $\psi_3=0$ in the action (\ref{SUSY WZW Action}). In the first-order formalism, we must also shift $\partial \bar W \to \partial \bar W+  A$:
\begin{align}
    S(A, \bar A) =& \frac{(k+2)}{2\pi}  \int d^2z \Big[\D r \dbar r + (\D \xi+A)(\dbar \xi + \bar A ) \nonumber \\  &\;\;\;\;\;\;\;\;\;\;\;\;\;\;\;\;\;\;\;\;\;\;\; + \chi (\dbar W + \bar A) + \bar \chi (\D \bar W + A) -\frac 1 {\sinh^2(r)}\chi \bar \chi \Big] \nonumber\\&\;\;\;\,+ \frac 1 {2\pi}\int d^2z\left[\psi_+(\dbar + \bar A)\psi_- +\ptild (\D +A)\mtild \right] \nonumber \\=&
    \,S_{\text{WZW}}|_{\psi_3=0}+ \frac {1} {2\pi} \int d^2z (J_3 \bar A + \bar J_3 A+ (k+2) A\bar A),
\end{align}
where 
\begin{equation}\label{current}
    J_3 = (k+2)(\partial \xi + \chi )+ {\p\m}, \;\; \bar J_3 = (k+2)(\dbar \xi + \bar \chi)+\ptild\mtild
\end{equation}
is the $U(1)$ current we would like to gauge. Integrating out $A$ and $\chi$ yields
\begin{align}
    S_\text{cigar}= \frac{(k+2)}{2\pi} &\int d^2 z [\D r \dbar r + \tanh^2(r) \D \theta \dbar \theta ] \nonumber\\ + \frac{1}{2\pi} &\int d^2z \,  \Big(\psi_+ (\dbar-\dbar\xi-i\tanh^2(r)\dbar \theta) \psi_- \nonumber \\&\;\;\;\;\;\;\;\;\;\;\;\;\;+\tilde\psi_+(\D-\D\xi+i\tanh^2(r)\D \theta  ) \mtild  \nonumber\\ 
    &
    \;\;\;\;\;\;\;\;\;\;\;\;\;-\frac {1}{(k+2)\cosh^2(r)}\p\m\ptild\mtild \Big),
\end{align}
which indeed is the supersymmetric sigma model into the Euclidean cigar with metric
\begin{equation}
    ds^2 = k(dr^2 + \tanh^2(r) d\theta^2). 
\end{equation}
This geometry represents a 2D Euclidean black hole \cite{Witten:1991yr, DIJGRAAF1992269}. In the asymptotic region $r\to \infty$, it approaches a cylinder with radius $\sqrt{k}$. The geometry ends at $r=0$, which is the tip of the cigar.

The fermions $\psi_a$ are no longer free after gauging. In the limit $r\to \infty$, however, we can define 
\begin{align}\label{new fermions}
    \lambda_\pm &= e^{\pm\xi\pm i\theta} \psi_\pm = e^{\pm \gamma/\sqrt{k+2}}\psi_\pm  \nonumber\\ \tilde \lambda_\pm &= e^{\pm \xi \mp i \theta} \tilde \psi_\pm = e^{\pm \bar \gamma /\sqrt{k+2}}\tilde \psi_\pm, 
\end{align}
which classically decouples the fermions. Due to the same chiral anomaly as the super-WZW model, quantum mechanically this change of variables shifts the coefficient of $\D\theta \dbar \theta$ back from $k+2\to k$. Defining the canonically normalized scalars $\rho = \sqrt{k}r, \phi = \sqrt{k} \theta$ and taking $\rho \to \infty$, the cigar model approaches the supersymmetric cylinder,
\begin{align}\label{cyl1}
    S_{\text{cigar}}\mathrel{\underset{(\rho\to \infty)}{\Longrightarrow}} S_{\text{cyl}}= \frac{1}{2\pi} &\int d^2 z \left[\D \rho \dbar \rho +  \D \phi \dbar \phi +\lambda_+\dbar \lambda_-  + \tilde \lambda _+ \D \tilde \lambda _-   \right].
\end{align}
Again, the dilaton is asymptotically linear, yielding a background charge $Q=1/\sqrt{k}$ for $\rho$. Moreover, $\phi \sim \phi+2\pi\sqrt{k}$, since the cylinder has radius $\sqrt{k}$. 

\subsection{$\mathcal{N}=2$ super-Liouville Theory}

The mirror dual to the Kazama-Suzuki coset theory above is $\mathcal{N}=2$ super-Liouville theory \cite{Ivanov:1983wp, SUSY_FZZ, NAKAYAMA_2004},
\begin{align}
    S_L= \frac 1 {2\pi} \int d^2 z  \bigg[ &\D \rho \dbar \rho + \D \phi \dbar \phi + \lp \dbar \lm + \lptild\D\lmtild  \nonumber \\&+2\pi \Big (\mu \lp \lptild e^{-\sqrt{k}\rho} e^{-i\sqrt{k}(\phi_L- \phi_R)} +\bar \mu \lm \lmtild e^{-\sqrt{k}\rho} e^{i\sqrt{k}(\phi_L - \phi_R)}\Big) \nonumber \\&+\frac{2\pi^2 |\mu|^2}{k} \normalorder{e^{-\sqrt{k}(\rho+i(\phi_L-\phi_R))}}\normalorder{e^{-\sqrt{k}(\rho-i(\phi_L-\phi_R))}} \bigg],
\end{align}
where $\phi \sim \phi+ 2\pi \sqrt{k}$. We also have a linear dilaton $\Phi(\rho) = -\rho/\sqrt{k}$. The last term is a contact term whose role is to regulate certain divergences. Amplitudes can be obtained by dropping this term and analytically continuing in momentum \cite{GREEN1988559, Hosomichi_2006}.

Taking $\rho\to \infty$, this too approaches the same free cylindrical theory,
\begin{equation}\label{cyl2}
    S = \frac{1}{2\pi  } \int d^2 z \left[\D \rho \dbar \rho + \D \phi \dbar \phi  +\lambda_+\dbar \lambda_-  + \tilde \lambda _+ \D \tilde \lambda_- \right]. 
\end{equation}
Dropping the contact term, the Liouville theory can be seen as a deformation of this free theory by the $(1, 1)$ potential
\begin{align}\label{liouville potential}
    V_L= 2\pi \left(\mu \lp \lptild e^{-\sqrt{k}\rho} e^{-i\sqrt{k}(\phi_L-\phi_R)} +\bar \mu \lm \lmtild e^{-\sqrt{k}\rho} e^{i\sqrt{k}(\phi_L-\phi_R)}\right).
\end{align}

As we have seen, both the Kazama-Suzuki supercoset and $\mathcal{N}=2$ Liouville theory asymptotically approach free cylindrical theories with a linear dilaton $\Phi \sim -\rho/\sqrt{k}$ which grows as we probe further into the geometry $\rho \to -\infty$. In the supercoset, the strong coupling region which would exist at $\rho = -\infty$ is cut off due to the geometry, which ends at the cigar tip $\rho=0$. On the other hand, in the Liouville theory, the geometry is still an infinite cylinder  $\rho \in (-\infty, \infty)$, but the potential $V_L$ exponentially suppresses the propagation of strings into the strong coupling region. The statement of the supersymmetric FZZ duality is that these CFTs are exactly equivalent to each other. It can be proven using mirror symmetry, as was done in \cite{SUSY_FZZ}.

\section{Uplifting the duality}\label{uplift}

As $\rho \to \infty$, the SUSY $\coset$ WZW model approaches the following free SCFT:
\begin{align}\label{free SCFT}
    S_{\text{free}}=\frac{1}{2\pi}&\int d^2z\left[\D \rho \dbar \rho + \beta  \dbar \gamma +\bar \beta \D \bar \gamma \right] \nonumber \\ +\frac{1}{2\pi} &\int d^2z \,  \left(\psi_+\dbar \psi_-+\tilde\psi_+\D \mtild + \frac 1 2(\psi_3\dbar \psi_3+\tilde \psi_3\D \tilde \psi_3 )\right),
\end{align}
where $\gamma \sim \gamma+2\pi i\sqrt{k+2}$, and $\rho$ has background charge $Q=1/\sqrt{k}$. By analogy with the 2D FZZ duality, we would like to deform this free SCFT by some potential $\V = \mu \V_+ + \bar \mu \V_-$ in order to extend the range of $\rho$ to $(-\infty, \infty)$.

The bosonic version of this duality has already been found in \cite{ER_EPR, Halder_2023}. For the level $k_B$ bosonic WZW model, the dual potential $W = \mu W_++ \bar \mu W_-$ is given by
\begin{equation}
    W^{(k_B)}_\pm (z, \bar z) = e^{-\sqrt{k_B-2}\rho }e^{\pm \frac{\sqrt{k_B}}4 (\gamma + \bar \gamma)}e^{\mp \sqrt{k_B}(\int^z \beta (z')dz' +\int^{\bar z} \bar \beta (z)dz)}.
\end{equation}
The supersymmetric duality follows from the bosonic case due to the decoupling of the fermionic sector. We simply dualize the bosonic part of the action in (\ref{SUSY WZW Action}) setting $k_B=k+2$.

We now verify that the operators $\V_\pm :=W^{(k+2)}_\pm$ reduce to the $\mathcal{N}=2$ Liouville potential upon gauging. The theory (\ref{free SCFT}) simply consists of the free boson $\rho$ with background charge $Q$, the cylindrical $\beta \gamma$ system, and the free fermions. The nontrivial OPEs are 
\begin{align*}
   \rho (z, \bar z)\rho(0, 0) &\sim -\frac 1 2\ln|z|^2 \\ \gamma (z) \beta (0) &\sim \frac 1 z \\ 
   \psi_+(z)\psi_-(0)&\sim \frac 1 z \\ \psi_3 (z) \psi_3(0) & \sim \frac 1 z ,
\end{align*}
and correspondingly for the antiholomorphic operators. The energy-momentum tensor is \begin{equation}
    T_B(z) = - (\D \rho)^2 + \sqrt{ \frac 1 k } \D^2 \rho   - \beta \D \gamma - \frac 1 2 ( \psi_3 \D \psi_3 + \p \D \m + \m \D \p ).
\end{equation}
The generator of the $U(1)$ we would like to gauge is the current (\ref{current}). In the normalized variables,\footnote{We drop terms $\D\bar \gamma = \dbar \gamma = 0$.}
\begin{align}\label{normalizedcurrent}
    J_3 &= \sqrt{k+2}\left (\beta + \frac 1 4  \D\gamma \right)+ \normalorder{\p\m}.  \nonumber \\  \bar J_3 & = \sqrt{k+2}\left( \bar \beta + \frac 1 4 \dbar \bar \gamma \right) + \normalorder{\ptild \mtild}.
\end{align}

Winding operators of the cylindrical $\beta \gamma$ system with periodicity $\gamma \sim \gamma + 2\pi i R$ take the form \cite{ER_EPR, Halder_2023, Frenkel:2005ku}
\begin{equation}
    \mathcal{O}_w(z, \bar z) = e^{-wR( \int^z dz'\beta(z')  + \int ^{\bar z} d \bar z ' \bar \beta (z'))}, 
\end{equation}
as can be seen from the OPE 
\begin{equation}\label{winding OPE}
   \gamma (z) \mathcal{O}_w(0) \sim w R \ln(z) \mathcal{O}_w(0).
\end{equation}
In our case, $R = \sqrt{k+2}$.

We will be interested in the operators with $w = \pm1$. By themselves, these operators are not states of the coset. They have a nonzero OPE with $j_3^B$, where $j_3^B$ is the purely bosonic current $j_3^B=\sqrt{k+2}\left (\beta + \frac 1 4  \D\gamma \right)$ \cite{ER_EPR}:
\begin{equation}
    j_3^B (z) \mathcal{O}_{\pm 1}(0) \sim \frac{\sqrt{k+2}}{4z} \mathcal{O}_{\pm 1 }(0).
\end{equation}
The OPE can be canceled by appending the operator $e^{\pm\frac{\sqrt{k+2}}4(\gamma+\bar \gamma)}$, leaving a primary with conformal weights $h = \bar h = (k+2)/4$; moreover, the winding OPE (\ref{winding OPE}) is unchanged. To obtain a primary with weight $(1, 1)$, we add the term $e^{-\sqrt{k}\rho}$, giving us the final operators \cite{ER_EPR, Halder_2023} 
\begin{equation}
    \V_\pm = e^{-\sqrt{k}\rho} e^{\pm \frac{\sqrt{k+2}}{4} (\gamma + \bar \gamma)}e^{\mp \sqrt{k+2} (\int^{z, \bar z} \beta (z')dz' + \bar \beta (\bar z') d\bar z')}.
\end{equation}

In the supersymmetric theory, the full $U(1)$ generator is simply the sum $J_3 = j_3^B + \normalorder{\p\m}$. Since the fermions are decoupled from the bosonic fields, the operators $\V_\pm$ are still states of the supersymmetric coset: $J_3(z)\V_\pm(0) = (j_3^B(z)+ \normalorder{\p\m(z)})\V_\pm(0) \sim 0$. In addition, since we have merely replaced the bosonic part of the theory with an equivalent CFT, the theory must still have $\mathcal{N}=1$ SUSY. The explicit supercurrent can be found through the bosonization introduced below. In terms of the variables $(\beta, \gamma, \psi_\pm)$, it has a rather complicated form:
\begin{align}
    T_F(z) = i \sqrt{2} \Bigg[ & \psi _3 \left(-\sqrt{ 1+ \frac 2 k }(\beta + \frac 1 4 \D \gamma ) - \sqrt{ \frac 1 k } \normalorder{\p\m}  \right)\nonumber \\&+ \frac 1 {\sqrt{2}} \normalorder{\p e^{\gamma /\sqrt{k+2}}}\left (\D \rho + \sqrt{1+ \frac 2 k }(-\beta+ \frac 1 4 \D \gamma ) -  \sqrt{ \frac 1 k } \normalorder{\p\m} - \frac{\D \gamma}{\sqrt{k(k+2)}} \right )\nonumber \\&+ \frac 1 {\sqrt{2}} \normalorder{\m e^{-\gamma /\sqrt{k+2}}}\left (\D \rho - \sqrt{1+ \frac 2 k }(-\beta+ \frac 1 4 \D \gamma ) +  \sqrt{ \frac 1 k } \normalorder{\p\m} + \frac{\D\gamma}{\sqrt{k(k+2)}}\right ) \nonumber \\&+ \sqrt{ \frac 1 {2k}}\D \left( \p e^{\gamma/\sqrt{k+2}} + \m e^{-\gamma/\sqrt{k+2}}\right) \Bigg ].
\end{align}
The operators $\V_\pm$ commute with this supercurrent; this is significantly easier to see with the bosonization below. 

To see how $\V$ reduces to the Liouville operator upon gauging, we can follow the strategy of \cite{Giveon_2003, Maldacena:2005hi}. We introduce a scalar $X_3$ to bosonize the current $J_3$,
\begin{equation}\label{current bosonization}
    J_3  = \sqrt{k+2}\left (\beta + \frac 1 4  \D\gamma \right)+\normalorder{\p \m}\simeq - \sqrt{{k}}\D X_3,
\end{equation}
and we also bosonize the fermions,
\begin{align}
    \normalorder{\psi _+ \psi _-} &\simeq i\sqrt{2}\D H.
\end{align}
We can now introduce an unnormalized scalar $Y$ defined by
\begin{equation}\label{Y def}
    \D Y = \sqrt{k+2}(-\beta + \frac 1 4 \D \gamma),
\end{equation}
which is orthogonal to $X_3$:
\begin{equation}
    \D X_3(z_1) \D Y(z_2) \propto  \beta (z_1) \partial \gamma (z_2) - \partial \gamma (z_1) \beta (z_2 )  \sim  -\frac{1}{z_{12}^2} +\frac 1{z_{12}^2}= 0.
\end{equation}

The scalar $Y$ can be split into two orthogonal, canonically normalized scalars $\phi_L$ and $H'$:
\begin{equation}\label{left Y}
    Y =- i\sqrt{k} \phi _L + i\sqrt{2} H'.
\end{equation}
$H'$ does not correspond to the bosonized fermions above. Instead, we have 
\begin{equation}
    i\sqrt{2}H' =  i\sqrt{2} H + \frac{\gamma}{\sqrt{k+2}}.
\end{equation}
One can verify that $\phi_L(z) H'(0) \sim 0$. By contrast, $H$ is not orthogonal to $\phi_L$. The need for this modification can be seen from (\ref{new fermions}): the fermions obtained from $H$ (i.e. $\psi_\pm$) are no longer free after gauging, and it is instead the fermions $\lambda_\pm = e^{\pm i \sqrt{2}H'}$ obtained by refermionizing $H'$ which are free in the coset. Note also that $J_3H'\sim 0$, and therefore $J_3 \phi_L \sim 0$ as well. It is straightforward to show that the energy-momentum tensor is also correctly mapped under the bosonization, $T_B = - \beta \D \gamma - (\D H)^2 = -(\D X_3)^2- (\D \phi)^2 - (\D H')^2 $. In these variables, the supercurrent takes a significantly simpler form, 
\begin{equation}
    T_F(z) = i \sqrt{2} \left [ \psi_3 \D X_3 +  \frac 1 {\sqrt{2}}\left(\lp \D (\rho -i\phi) + \lm \D (\rho + i \phi) + \sqrt{\frac 1 k} \D (\lp + \lm) \right)\right].
\end{equation}

One can also repeat this entire procedure for the left-moving sector, defining\footnote{The alternate choice of signs would give us the T-dual Liouville theory.}
\begin{equation}\label{right Y}
   \dbar  \tilde Y  = \sqrt{k+2} ( -\bar  \beta + \frac 1 4 \dbar \bar \gamma )  =i\sqrt{k} \dbar \phi_R + i\sqrt{2}\dbar\tilde H'.
\end{equation}
Now, note that 
\begin{equation}
    \V _\pm= e^{-\sqrt{k}\rho}e^{\pm\int \partial Y dz'+ \dbar \tilde Yd \bar z'} = e^{-\sqrt{k}\rho}e^{\pm (Y + \tilde Y)}. 
\end{equation}
Written as such, it is clear that $J_3 \V \sim 0$, since $X_3$ and $Y$ are orthogonal. Thus, the operator $\V$ is unchanged upon gauging. Substituting the expressions (\ref{left Y}) and (\ref{right Y}), we see that the operator $\V$ takes the form of the $\mathcal{N}=2$ Liouville operator,
\begin{equation}
    \V_\pm = e^{-\sqrt{k}\rho}e^{\pm i\sqrt{2}(H'+ \tilde H') \mp i \sqrt{k} (\phi_L - \phi_R)} \simeq \lambda_\pm \tilde \lambda _\pm e^{-\sqrt{k}\rho}e^{\mp i \sqrt{k}(\phi_L - \phi_R) },
\end{equation}
where we have refermionized the expressions $e^{\pm i\sqrt{2}H'}\simeq \lambda_\pm, e^{\pm i \sqrt{2}\tilde H'}\simeq \tilde \lambda_\pm$. 

To see that the periodicity is also correct, we note that setting $X_3=\bar X_3 = 0$ gives the conditions
\begin{align*}
    \int \beta dz &= - \frac{i\sqrt{2}}{\sqrt{k+2}} H - \frac \gamma 4, \\\int \bar \beta d \bar z &= - \frac{i\sqrt{2}}{\sqrt{k+2}} \tilde H - \frac {\bar  \gamma} 4.
\end{align*}
Substituting into (\ref{left Y}) and (\ref{right Y}), we find 
\begin{equation}
    \phi = \phi_L + \phi_R =  \frac i 2 \sqrt{\frac k {k+2}}  (\gamma - \bar \gamma).
\end{equation}
Under $\gamma \to \gamma + 2\pi i \sqrt{k+2}$, we have $\phi \to \phi - 2\pi \sqrt{k}$. Hence, $\phi \sim \phi + 2\pi \sqrt{k}$, which is the periodicity of the Liouville theory. 

In sum, after gauging the current $J_3$ given by (\ref{normalizedcurrent}) and projecting $\psi _3 = 0$, the CFT 
\begin{align}
    S=\frac{1}{2\pi}&\int d^2z\left[\D \rho \dbar \rho + \beta  \dbar \gamma +\bar \beta \D \bar \gamma + 2\pi (\mu \V_+ + \bar \mu \V_-)\right] \nonumber \\ +\frac{1}{2\pi} &\int d^2z \,  \left(\psi_+\dbar \psi_-+\tilde\psi_+\D \mtild + \frac 1 2(\psi_3\dbar \psi_3+\tilde \psi_3\D \tilde \psi_3 )\right)
\end{align}
reduces to $\N = 2$ Liouville theory.

\section{Conclusion}

In this short note, we have shown that the supersymmetric version of the AdS$_3$ duality of \cite{ER_EPR, Halder_2023} follows trivially from the bosonic case due to the decoupling of the fermionic sector in the $H_3^+$ super-WZW model. To verify this, we have shown that the winding potential of the dual CFT reduces to the $\mathcal N = 2$ Liouville potential upon gauging the timelike isometry $J_3$. 

Due to the somewhat unusual decoupling of the fermions in the supersymmetric theory, one could potentially evaluate the partition function of the dual on various manifolds by exploiting the localization arguments of \cite{murthy2025localizationstringsgroupmanifolds}. The partition function of $\coset$ diverges, so one may need to analytically continue or quotient the CFT first. 

It would be interesting to find a proof of this duality using a gauged linear sigma model, as was done for the original supersymmetric FZZ duality in \cite{SUSY_FZZ}. Such a linear sigma model may provide insight into the black hole/string transition in AdS$_3$ \cite{Chen:2021dsw, Urbach_2022, Urbach:2023npi, agia2023ads3stringstarspure}.

\section*{Acknowledgments}

I am grateful to Gabriel Wong and James Read for discussions and feedback.

\bibliographystyle{JHEP}
\bibliography{refs.bib}

\end{document}